\documentclass[conference,11pt]{IEEEtran}

\usepackage[utf8]{inputenc}
\usepackage[T1]{fontenc}
\usepackage{graphicx}
\usepackage{amsmath,amssymb,amsfonts}
\usepackage{booktabs}
\usepackage{xcolor}
\usepackage{url}
\usepackage{cite}
\usepackage{etoolbox}
\usepackage{microtype}
\usepackage{placeins}
\usepackage{enumitem}
\setlist{nosep, leftmargin=14pt}

\usepackage{caption}
\usepackage{subcaption}

\usepackage{hyperref}

\hypersetup{
  colorlinks=true,
  linkcolor=black,
  citecolor=black,
  urlcolor=black,
}

\newcommand{\best}[1]{\textbf{\textit{#1}}}

\begin{document}

\title{Deep Unfolded BM3D: Unrolling Non-local Collaborative Filtering into a Trainable Neural Network}

\author{
\IEEEauthorblockN{Kerem Basim\textsuperscript{1}, Mehmet Ozan Unal\textsuperscript{1}, Metin Ertas\textsuperscript{2}, Isa Yildirim\textsuperscript{1}}
\\
\IEEEauthorblockA{\textsuperscript{1}Electronics and Communication Engineering Department, Istanbul Technical University, Istanbul, Turkey\\
\textsuperscript{2}Electrical and Electronics Engineering Department, Istanbul University, Istanbul, Turkey\\
Email: basim21@itu.edu.tr, unalmehmet@itu.edu.tr, ertas@istanbul.edu.tr, iyildirim@itu.edu.tr}
}

\maketitle

\begin{abstract}
Block-Matching and 3D Filtering (BM3D) exploits non-local self-similarity priors for denoising but relies on fixed parameters. Deep models such as U-Net are more flexible but often lack interpretability and fail to generalize across noise regimes. In this study, we propose Deep Unfolded BM3D (DU-BM3D), a hybrid framework that unrolls BM3D into a trainable architecture by replacing its fixed collaborative filtering with a learnable U-Net denoiser. This preserves BM3D's non-local structural prior while enabling end-to-end optimization. We evaluate DU-BM3D on Low-Dose CT (LDCT) denoising and show that it outperforms classic BM3D and standalone U-Net across simulated LDCT at different noise levels, yielding higher PSNR and SSIM, especially in high-noise conditions.
\end{abstract}


\section{INTRODUCTION}

X-ray Computed Tomography (CT) is essential in clinical practice but faces a fundamental trade-off between image quality and radiation exposure. Low-Dose CT (LDCT) reduces patient risk but introduces strong noise that obscures anatomy and degrades diagnostic confidence. Post-processing denoising is flexible and does not depend on specific scanner models, but generalizing across dose levels without protocol-specific retraining is challenging because CT acquisition settings vary by anatomy, patient, and task.

Traditional model-based methods like BM3D~\cite{dabov2007image} exploit non-local self-similarity but rely on fixed parameters. Deep learning approaches such as DnCNN~\cite{zhang2017beyond} and GANs learn noise-to-clean mappings but lack interpretability and often require retraining per dose regime. In this study, we introduce a hybrid framework ``Deep Unfolded BM3D (DU-BM3D)'' that unrolls BM3D into a trainable network by replacing its fixed collaborative filtering with a learnable U-Net denoiser. This preserves BM3D's structural prior while enabling end-to-end optimization. We demonstrate that a single DU-BM3D model, trained at one dose level, generalizes across simulated LDCT noise levels ranging from 10k to 500k photon counts, outperforming both classical BM3D and standalone U-Net.

\begin{figure*}[t]
    \captionsetup{width=\textwidth,justification=centering}
    \centering
    \includegraphics[width=0.95\textwidth]{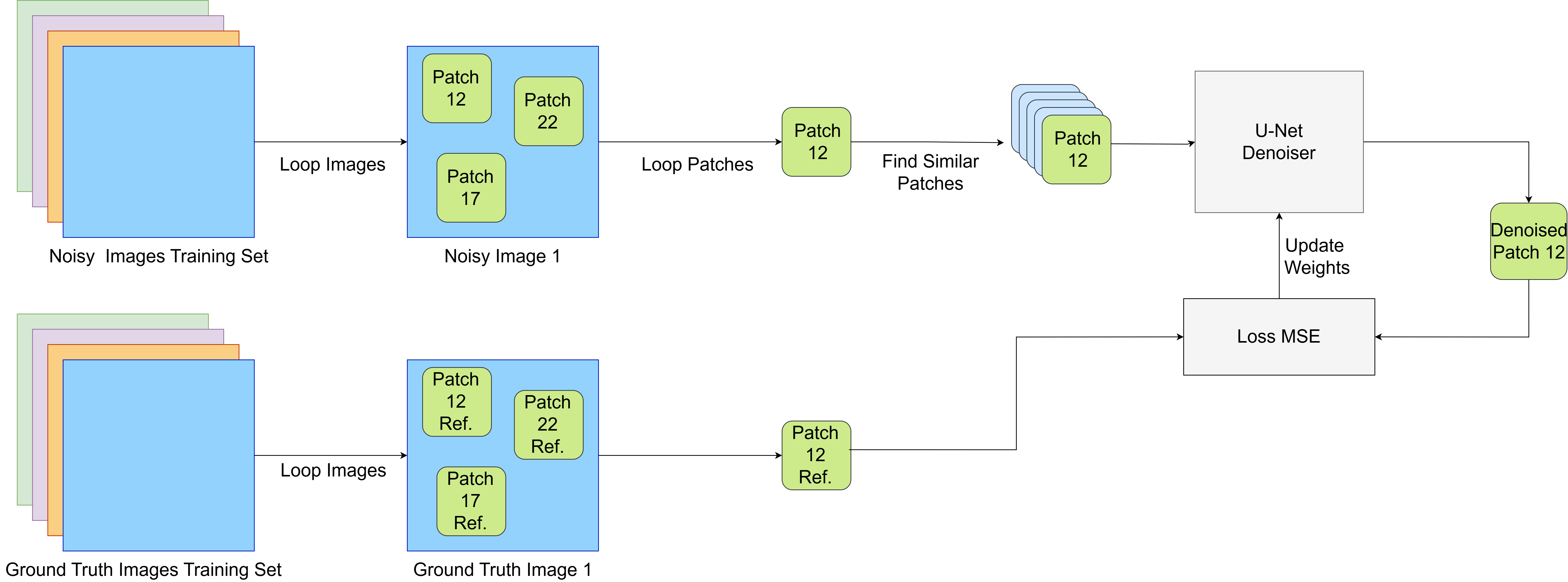}
    \caption{Training procedure for the U-Net denoiser component.}
    \label{fig:diagram}
\end{figure*}

\section{RELATED WORK}
\label{sec:related_work}

LDCT denoising has been pursued through (i) traditional model-based algorithms, (ii) data-driven deep learning, and (iii) hybrid deep unfolding approaches.

\noindent\textbf{Model-based denoising:}
Classical methods impose explicit image priors: Total Variation (TV) enforces edge-preserving smoothness, wavelet models use multi-scale sparsity, and Non-Local Means (NLM) averages self-similar patches. Block-Matching and 3D Filtering (BM3D)~\cite{dabov2007image} is particularly effective in medical imaging: it groups similar patches into 3D stacks, applies transforms (e.g., DCT), and attenuates noise via coefficient shrinkage. BM3D and its LDCT-oriented extensions (e.g., context-aware BM3D~\cite{chen2016denoising}) perform strongly, but rely on fixed hand-tuned parameters and adapt poorly across dose levels.

\noindent\textbf{Deep learning-based denoising:}
Deep learning methods learn direct mappings from low-dose to normal-dose CT. U-Net~\cite{jin2017deep} is widely used in biomedical imaging, via an encoder–decoder with skip connections that fuses semantic context with fine detail. Subsequent work incorporates self-attention~\cite{wang2018non} and non-local similarity modeling~\cite{song2024real}. However, such models act as ``black boxes'', lack explicit priors, and require large annotated datasets that are difficult to obtain in medical imaging.

\noindent\textbf{Hybrid / deep unfolding:}
Deep Unfolding (DU)~\cite{monga2021algorithm, mou2022deep} maps iterative optimization procedures to neural networks, exposing interpretable structure while allowing parameters to be learned. BM3D-Net~\cite{yang2017bm3d} follows this direction by learning transform-domain filters. Our approach differs in scope: we unfold the full BM3D pipeline and replace its entire filtering stage with a learnable U-Net denoiser, yielding an end-to-end trainable model that couples BM3D's non-local structural prior with data-driven adaptability.

\section{METHODOLOGY}
\label{sec:method}

In this study, we introduce a hybrid framework integrating model-based algorithms with deep learning. Inspired by Deep Unfolding (DU)~\cite{monga2021algorithm}, we re-interpret BM3D~\cite{dabov2007image} as a learnable network. BM3D comprises three stages: (1) block-matching, (2) collaborative filtering, and (3) aggregation. We keep the first and third stages as fixed operators, preserving their mathematical structure, and replace the second stage with a compact learnable U-Net~\cite{ronneberger2015u}, yielding a parameter-efficient architecture. Our proposed Deep Unfolded BM3D (DU-BM3D) framework consists of three steps:

\noindent\textbf{Block-matching (non-learnable):}
Given a noisy low-dose CT input $x_l \in \mathbb{R}^{H \times W}$, we apply a block-matching operator $\mathcal{M}(\cdot)$ that leverages non-local self-similarity: it searches for similar patches and groups them into 3D stacks $G_l$,
\begin{equation}
    G_l = \mathcal{M}(x_l)
    \label{eq:matching}
\end{equation}
with patch size, search window, and distance threshold fixed during training.

\noindent\textbf{Learnable collaborative filtering:}
We replace BM3D's fixed transform-domain filtering with a compact learnable denoiser $D_{\theta}(\cdot)$ implemented as a U-Net. It takes the noisy stack $G_l$ and outputs a denoised stack $\hat{G}_n$,
\begin{equation}
    \hat{G}_n = D_{\theta}(G_l)
    \label{eq:filtering}
\end{equation}
where the encoder–decoder with skip connections learns to separate signal from noise using both local and non-local correlations.

\begin{figure*}[t]
    \captionsetup{width=\textwidth,justification=centering}
    \centering

    \includegraphics[width=0.95\textwidth]{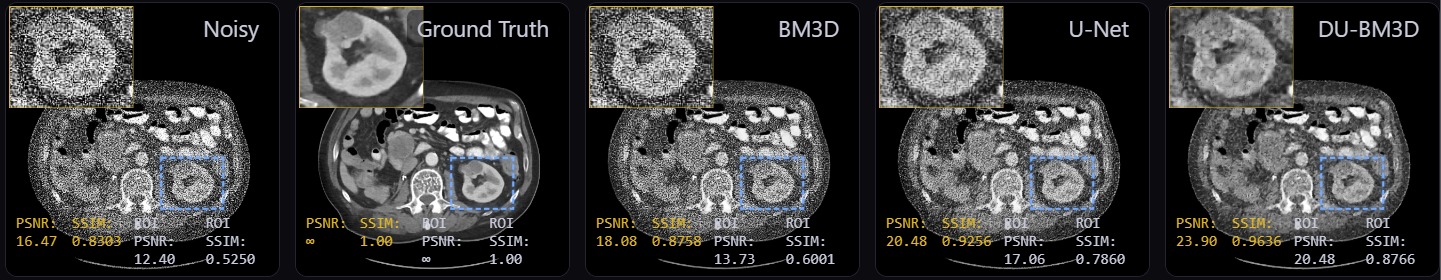}\\[0.5em]
    \includegraphics[width=0.95\textwidth]{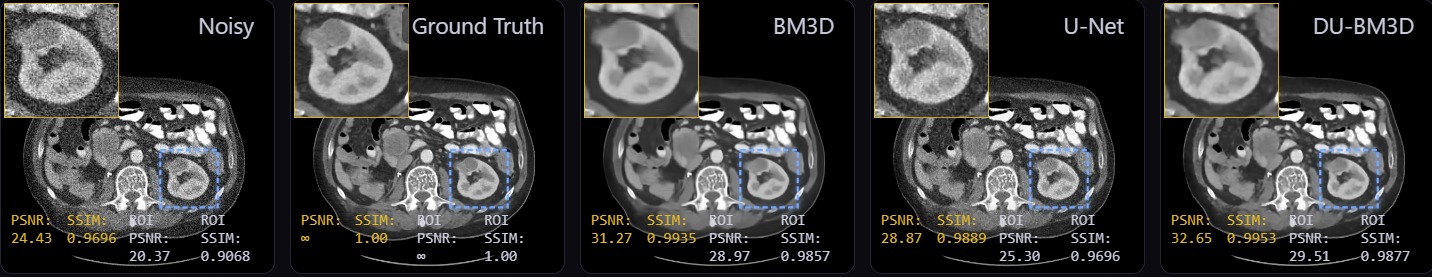}

    \caption{Denoising results at different photon levels. Top row: 10k photon
    level (extreme low-dose). Bottom row: 100k photon level (moderate
    low-dose). From left to right in each row: Noisy input, Ground Truth,
    BM3D, U-Net, and DU-BM3D (ours).}
    \label{fig:qualitative}
\end{figure*}

\noindent\textbf{Aggregation (non-learnable):}
The denoised stacks $\hat{G}_n$ are re-projected to their original 2D locations using a fixed aggregation operator $\mathcal{A}(\cdot)$, which reconstructs the final image $\hat{x}_n$ via weighted averaging of overlapping patches:
\begin{equation}
    \hat{x}_n = \mathcal{A}(\hat{G}_n)
    \label{eq:aggregation}
\end{equation}

The complete DU-BM3D model $f_{\theta}(\cdot)$ is the composition of these three stages \eqref{eq:matching}, \eqref{eq:filtering} and \eqref{eq:aggregation}:
\begin{equation}
  \hat{x}_n = f_{\theta}(x_l) = \mathcal{A}\!\big(D_{\theta}(\mathcal{M}(x_l))\big).
  \label{eq:fullmodel}
\end{equation}

For each training pair $(x_l, x_n)$ of low-dose and normal-dose CT images, overlapping patches are extracted from both $x_l$ and $x_n$. Similar patches from the noisy image are grouped via block-matching, forming non-local 3D stacks that are fed to the U-Net denoiser $D_{\theta}(\cdot)$. We optimize the U-Net parameters $\theta$ to minimize the discrepancy between the predicted output $\hat{x}_n$ and the ground truth $x_n$.

Given $N$ training pairs $\{x_{l,i}, x_{n,i}\}_{i=1}^{N}$, we minimize the Mean Squared Error (MSE) loss:
\begin{equation}
  \mathcal{L}(\theta) = \frac{1}{N} \sum_{i=1}^{N} \| f_{\theta}(x_{l,i}) - x_{n,i} \|_2^2 .
  \label{eq:loss}
\end{equation}

Optimization uses the Adam optimizer with backpropagation. Since $\mathcal{M}$ and $\mathcal{A}$ are fixed, gradients flow \emph{only} through the U-Net parameters $\theta$, forcing it to learn denoising \emph{within} BM3D's structured context. The overall training pipeline is illustrated in Fig.~\ref{fig:diagram}.

\section{EXPERIMENTS}
\label{sec:experiments}

We validate DU-BM3D through comparison with BM3D and U-Net baselines. We used CT images from DeepLesion~\cite{yan2018deeplesion}, following the LoDoPaB-CT simulation procedure~\cite{leuschner2021lodopab} to generate four dose levels (10k, 50k, 100k, 500k photon counts). All models were trained \emph{only} on 100k-photon data and then evaluated across the entire dose range (10k--500k) without retraining. This single-model evaluation protocol directly assesses cross-dose generalization, in contrast to dose-specific deep learning approaches that require a separate network per noise regime. The dataset was split into training, validation, and test subsets with a 69\%, 14\%, and 17\% distribution, respectively. All models were implemented in PyTorch and trained for 20 epochs using Adam and MSE loss, with batch size 16 on an NVIDIA A100 GPU. The code, trained DU-BM3D model, and evaluation scripts are available online.\footnote{https://github.com/itu-biai/deep\_unfolded\_bm3d}

We compared our proposed DU-BM3D against two baselines: (1) conventional BM3D~\cite{dabov2007image} using a fixed configuration without any dose-specific retuning, and (2) a standalone U-Net~\cite{jin2017deep} with the same architecture as our denoiser $D_{\theta}$, trained end-to-end on paired low-dose/normal-dose CT scans.

\begin{table}[t]
\centering
\caption{Average PSNR (dB) and SSIM at different photon levels.}
\label{tab:denoise_metrics_single}
\small
\setlength{\tabcolsep}{1.5pt}
\renewcommand{\arraystretch}{1.0}
\begin{tabular}{r cc cc cc cc}
\hline
Photon & \multicolumn{2}{c}{Noisy} & \multicolumn{2}{c}{BM3D} & \multicolumn{2}{c}{DU-BM3D} & \multicolumn{2}{c}{U-Net} \\
 & PSNR & SSIM & PSNR & SSIM & PSNR & SSIM & PSNR & SSIM \\
\hline
10k  & 15.63 & 0.6304 & 17.79 & 0.6638 & \best{24.15} & \best{0.7417} & 20.47 & 0.6887 \\
50k  & 20.67 & 0.7060 & 28.05 & 0.8214 & \best{30.34} & \best{0.8657} & 25.49 & 0.7751 \\
100k & 23.04 & 0.7423 & 29.87 & 0.8600 & \best{31.77} & \best{0.8848} & 27.71 & 0.8184 \\
500k & 28.07 & 0.8295 & 30.34 & 0.8653 & \best{31.99} & \best{0.8861} & 30.84 & 0.8891 \\
\hline
\end{tabular}
\end{table}

Table~\ref{tab:denoise_metrics_single} reports PSNR and SSIM across all dose levels. DU-BM3D consistently outperforms both classic BM3D and standalone U-Net, with the largest gains in the high-noise regime (10k photons), where it reaches 24.15 dB PSNR versus 20.47 dB for U-Net and 17.79 dB for BM3D. Similar trends hold at 50k, 100k, and 500k photons, indicating that combining BM3D's structural prior with a learned U-Net denoiser yields superior denoising quality. Importantly, a \emph{single} DU-BM3D model trained at one dose generalizes across all dose levels without retraining, outperforming dose-specific deep learning baselines.

Fig.~\ref{fig:qualitative} illustrates qualitative results at 10k and 100k photons with related PSNR and SSIM values together with a zoomed-in specific ROI. At 10k (top row), BM3D leaves structured artifacts, while U-Net over-smooths and blurs fine anatomy. DU-BM3D suppresses noise while preserving anatomical edges and lesion boundaries. At 100k (bottom row), DU-BM3D continues to preserve tissue boundaries and lesion characteristics that are diagnostically relevant.

\begin{figure}[t]
    \centering
    \subfloat[Model Complexity (Parameters)]{
        \includegraphics[width=0.97\columnwidth]{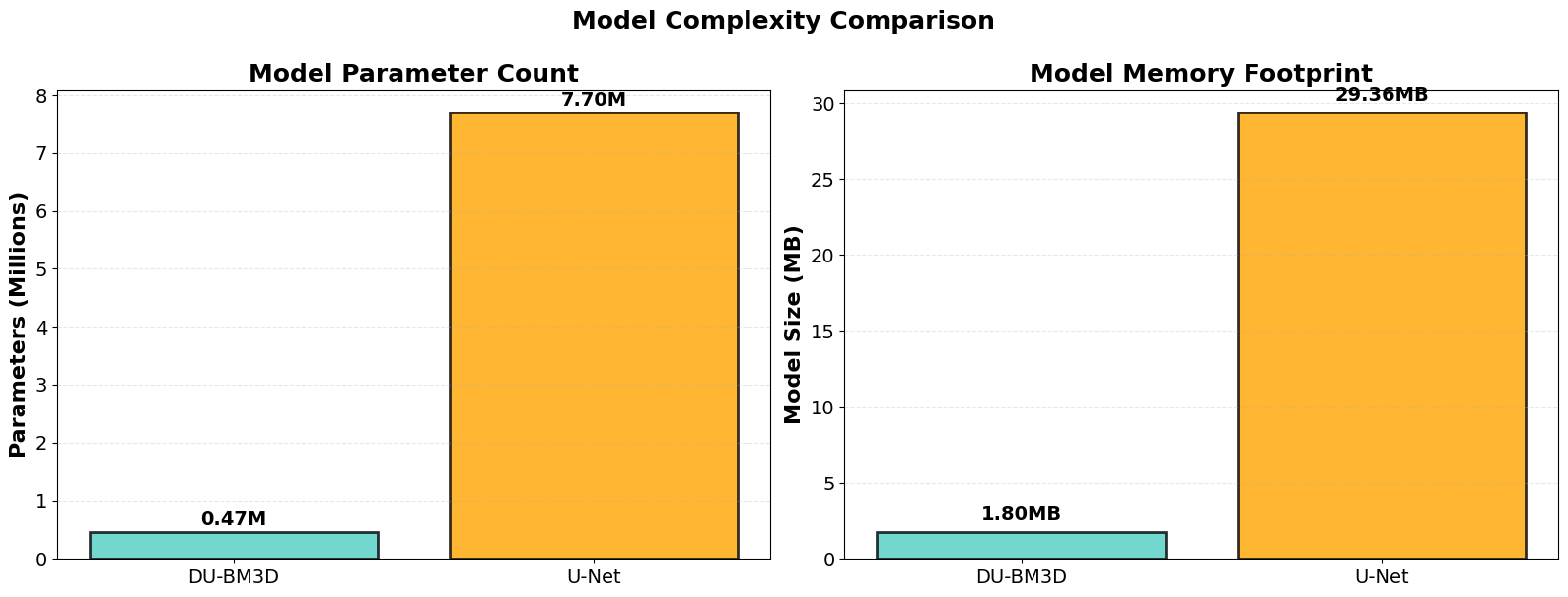}
        \label{fig:analysis_params}
    }\\[-0.25em]
    \subfloat[Average Inference Time]{
        \includegraphics[width=0.97\columnwidth]{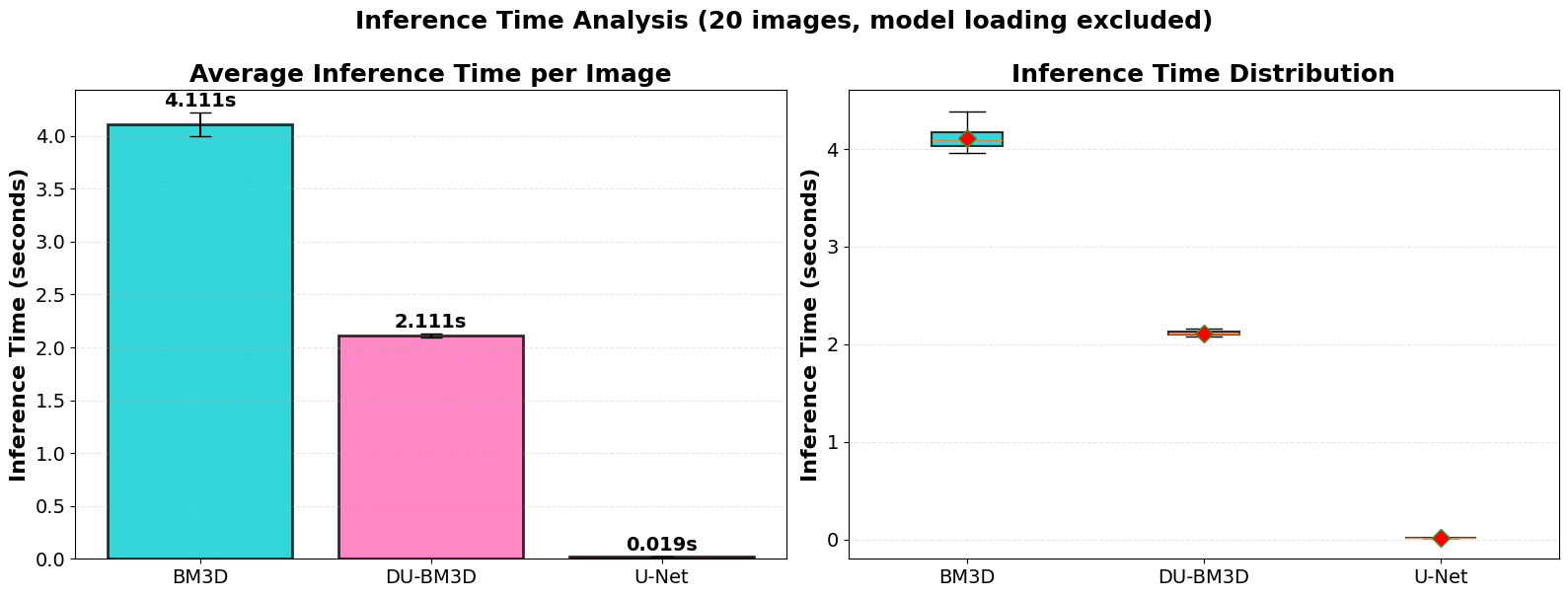}
        \label{fig:analysis_time}
    }
    \caption{Comparison of models regarding (a) number of learnable parameters and (b) average inference time per test image.}
    \label{fig:analysis}
\end{figure}

Fig.~\ref{fig:analysis} summarizes computational aspects. DU-BM3D uses much fewer parameters than the standalone U-Net (Fig.~\ref{fig:analysis_params}). Considering its inference time (Fig.~\ref{fig:analysis_time}), DU-BM3D is faster than BM3D and but moderately slower than U-Net. Overall, DU-BM3D provides a practical balance between denoising performance, parameter efficiency, and computational cost.

\section{DISCUSSION}
\label{sec:discussion}

Our results demonstrate that integrating data-driven learning into a successful model-based framework yields superior performance compared to either approach alone. BM3D is one of the most effective classical denoising methods in medical imaging. By making its collaborative filtering stage learnable via U-Net, we achieved significant improvements while maintaining interpretability. We chose U-Net for its established effectiveness in image-domain CT processing (e.g., FBP + U-Net), and used the same architecture for both patch-based and standalone implementations to ensure fair comparison.

Our approach represents one specific strategy for making BM3D learnable, replacing collaborative filtering while keeping block-matching and aggregation fixed. However, other BM3D components such as matching criteria, transform operations, and aggregation weights could also benefit from data-driven adaptation. Future work will investigate these alternative pathways for incorporating data-driven adaptation into BM3D.

\section{CONCLUSION}
\label{sec:conclusion}

In this work, we introduced Deep Unfolded BM3D (DU-BM3D), a hybrid framework for Low-Dose CT (LDCT) denoising that integrates the interpretable non-local self-similarity prior of BM3D with the adaptive learning capacity of deep neural networks. By preserving BM3D's non-local structural prior while introducing a learnable denoiser, DU-BM3D is designed to achieve robust cross-dose generalization, a critical requirement for clinical deployment, where radiation protocols are patient- and task-specific.

More broadly, this work supports the view that hybrid approaches which fuse domain-specific structural priors with data-driven learning offer a principled and practical direction for medical image restoration in diverse clinical scenarios.

\apptocmd{\thebibliography}{\setlength{\itemsep}{0pt}}{}{}
\bibliographystyle{IEEEtran}
\bibliography{refs}

\end{document}